\documentclass{article}
\usepackage{spconf,amsmath,graphicx}
\usepackage{hyperref}
\usepackage{booktabs}
\usepackage{multirow}
\usepackage{enumitem}
\usepackage{bibspacing}
\usepackage[numbers,sort&compress]{natbib}
\title{SEF-VC: Speaker Embedding Free Zero-Shot Voice Conversion with Cross Attention}
\name{Junjie Li, Yiwei Guo, Xie Chen, $^{\dagger}$Kai Yu
\thanks{$^{\dagger}$Kai Yu is the corresponding author.}}
\address{X-LANCE Lab, Department of Computer Science and Engineering\\MoE Key Lab of Artificial Intelligence, AI Institute\\Shanghai Jiao Tong University, Shanghai, China\\
\small{\texttt{
\{junjieli, yiwei.guo, chenxie95, kai.yu\}@sjtu.edu.cn
}}
}

\begin{document}
%
\maketitle
\begin{abstract}
Zero-shot voice conversion (VC) aims to transfer the source speaker timbre to arbitrary {\em unseen} target speaker timbre, while keeping the linguistic content unchanged.
Although the voice of generated speech can be controlled by providing the speaker embedding of the target speaker, the speaker similarity still lags behind the ground truth recordings.
In this paper, we propose SEF-VC, a speaker embedding free voice conversion model, 
which is designed to learn and incorporate speaker timbre from reference speech via a powerful position-agnostic cross-attention mechanism, and then reconstruct waveform from HuBERT semantic tokens in a non-autoregressive manner.
The concise design of SEF-VC enhances its training stability and voice conversion performance.
Objective and subjective evaluations demonstrate the superiority of SEF-VC to generate high-quality speech with better similarity to target reference than strong zero-shot VC baselines, even for very short reference speeches.
\end{abstract}
\section{Introduction}
\label{sec:intro}

Zero-shot voice conversion (VC) is tasked to convert the given speech from the source speaker to a previously unseen target speaker while preserving the speech content.
This mainly involves two difficulties: disentanglement of speaker and content information, and speaker representation modeling. 
Disentanglement aims to remove speaker information from source speech, while speaker representation modeling seeks a better way to represent and incorporate speaker identities.

There are consistent and active research efforts in the disentanglement of speaker information. VC methods based on auto-encoders are first developed, which learn meaningful latent representations by designing information bottlenecks in the speech reconstruction process~\cite{hsu2016voice,qian2019autovc,qian2020unsupervised,chan2022speechsplit2,wang2022drvc}.
These bottleneck features can disentangle speaker information to some extent, but usually with sacrificed speech quality~\cite{wang2021noisevc}.
Normalizing flows~\cite{kingma2018glow} then provide a more elegant way for speaker adaptation and VC~\cite{casanova2021sc,du2023speaker}, such as YourTTS~\cite{casanova2022yourtts}.
A more popular technique for speaker disentanglement recently is the adoption of self-supervised semantic features, such as
vq-wav2vec~\cite{baevski2019vq} and HuBERT~\cite{hsu2021hubert}.
Features extracted by these models are proved to well preserve the linguistic content while being scarcely speaker-variant \cite{polyak2021speech,du2023speaker}. 
Previous studies introduce self-supervised semantic features into VC, but still in the traditional auto-encoder framework \cite{lin2021s2vc,dang2022training}, or not in an any-to-any fashion~\cite{huang2021any}.
\cite{polyak2021speech} proposes to train a vocoder that synthesizes waveforms from semantic tokens, which also brings an easier VC framework.

However, the representation modeling of speaker identity in the process of copying the target voice still needs to be investigated.
Most VC methods rely on a global speaker embedding, especially from speaker verification networks~\cite{snyder2018x}.
\cite{qian2019autovc,saito2018non} employ a pretrained speaker encoder.
\cite{lian2022towards,lian2022robust} sample speaker embeddings from a posterior distribution. \cite{xiao2022dgc, tan2021zero} introduce a speaker representation method to better represent the characteristics of target speakers.
The VC performance of these methods is thus limited by the representation ability of speaker embeddings, and they are also not robust to short references.
Recent speech language models~\cite{borsos2023audiolm,wang2023lm,huang2023make} avoid this issue by the promising in-context learning strategy that predicts the target voice given speech prompts.
But they also suffer from stability issues due to their autoregressive nature.


Different from previous works, we propose SEF-VC, a speaker embedding free zero-shot VC model.
We propose to use the position-agnostic cross-attention mechanism~\cite{du2023unicats} as the speaker modeling approach.
This replaces the conventional speaker embedding approach with a novel, effective, and robust \textbf{cross-attention} mechanism~\cite{du2023unicats}.
Then, SEF-VC is designed to learn and incorporate speaker timbre from reference speech via this type of cross-attention, and then reconstruct waveform from HuBERT semantic tokens in a non-autoregressive manner. 
With the position-agnostic cross-attention mechanism, the speaker information is better modeled and incorporated into the semantic backbone.
Objective and subjective evaluations demonstrate that SEF-VC outperforms several strong VC baselines.
Ablation studies further show the advantage of using cross-attention instead of speaker embeddings, and the competence of SEF-VC over different prompt lengths.
Audio samples are available at \url{https:///junjiell.github.io/SEF-VC/}.

\vspace{-0.5cm}
\section{SEF-VC}
\label{sec:method}
\label{sec:model architecture}
\begin{figure}[ht]
    \centering
    \includegraphics[width=0.99\linewidth]{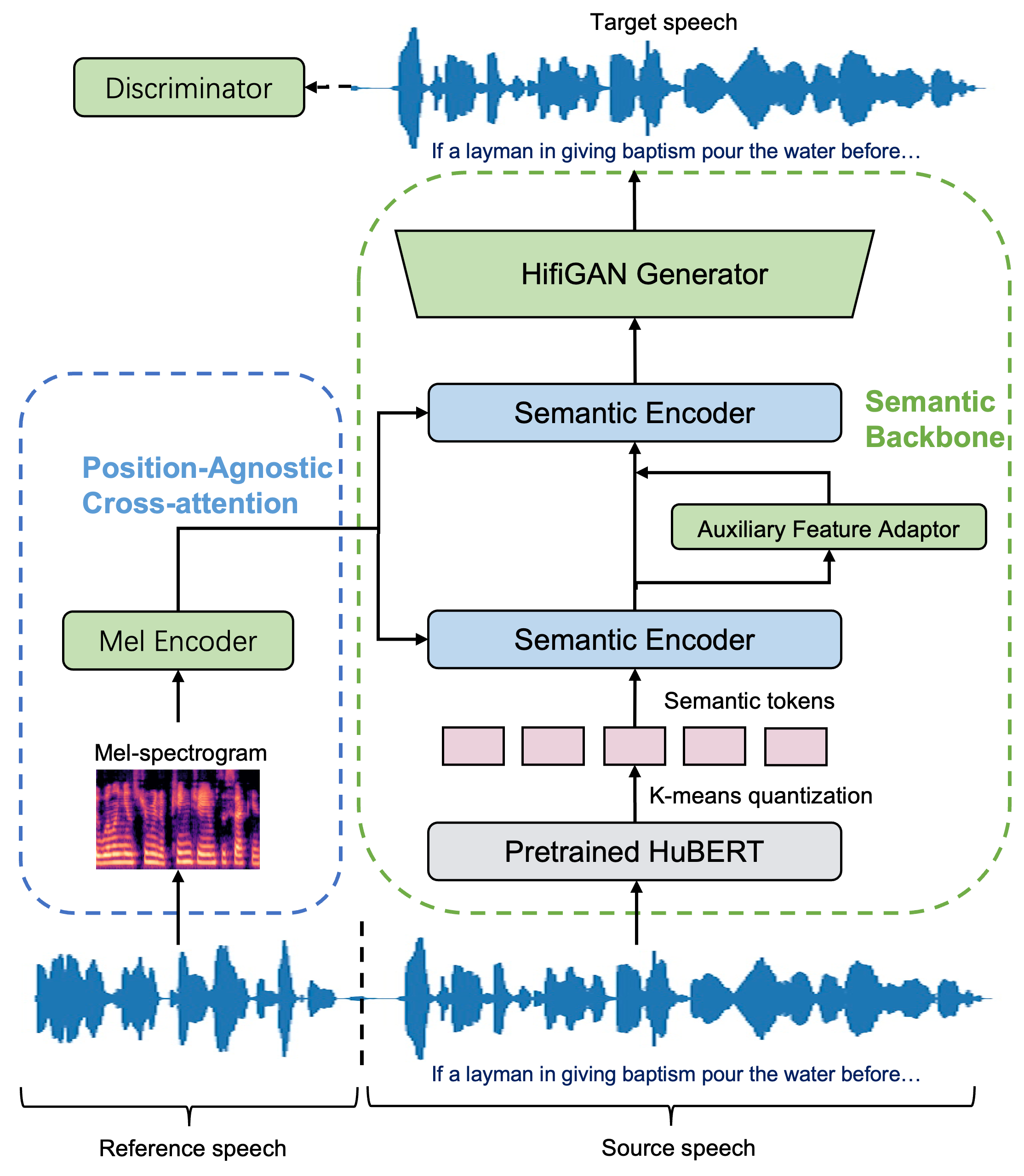}
    \caption{The model architecture of SEF-VC.}
    \label{fig:model}
\end{figure}

\vspace{-0.5cm}
\subsection{Non-Autoregressive Semantic Backbone}
The architecture of the model is illustrated in Fig.~\ref{fig:model}. 
The backbone model of SEF-VC is a standard vocoder on discrete self-supervised speech representations, like vec2wav~\cite{ du2023unicats} and SSR-VC~\cite{polyak2021speech}.
We first obtain semantic tokens by conducting K-Means quantization on the continuous features extracted by a pretrained HuBERT model.
Then, the frame-level semantic tokens are passed through two semantic encoders, and then upsampled into waveforms in a HifiGAN generator. 
Following \cite{du2023unicats}, an auxiliary feature adaptor is placed between two semantic encoders, which helps to model the speech prosody features by predicting the pitch, probability of voice, and energy (PPE).

To enhance the synthesis quality, we also adopt adversarial training following the practices in vocoders~\cite{kong2020hifi,du2023unicats}.
The discriminator here comprises a multi-period discriminator (MPD) and a multi-scale discriminator (MSD) proposed in HifiGAN~\cite{kong2020hifi}, which tries to distinguish reconstructed waveforms against ground truth ones.
\vspace{-0.4cm}
\subsection{Position-Agnostic Cross-Attention Mechanism}
As the self-supervised semantic tokens scarcely provide speaker information, the conversion of timbre must be dependent on an explicit introduction of speaker information.
Rather than previous works that adopt speaker embeddings, we formulate this as a speaker embedding free cross-attention task, where the semantic backbone learns and incorporates speaker timbre directly from the reference speech.
To introduce the reference speech with sufficient speaker information from which we extract mel-spectrograms, we use the position-agnostic cross-attention mechanism to incorporate speaker information into the semantic backbone. 

Specifically, in the semantic backbone, each of 
the semantic encoders consist of several Conformer blocks~\cite{conformer}.
Within each conformer block, we place a cross-attention layer between the self-attention layer and the convolution module.
Before cross-attention, the mel-spectrogram of the target speech is fed through a mel encoder which consists of 
a 1D-convolution layers as a pre-net.
Notably, this cross-attention mechanism is agnostic of input positions, which means the positional encodings in the standard attention mechanism are canceled when computing key and value matrices from encoded mel sequence.
This is equivalent to shuffling the encoded mel-spectrograms, and since speaker timbre is mostly irrelevant to time order, breaking the sequential order will still maintain a considerable amount of speaker information but not others.
This helps the cross-attention mechanism focus on learning to capture speaker timbre from the reference speech.

This cross-attention mechanism is also beneficial for both short and long references.
For short references, the cross-attention mechanism can still fully explore and utilize the mel-spectrograms directly, without the risk of inaccurate speaker modeling from speaker embeddings.
For long references, the position-agnostic cross-attention mechanism provides an additional advantage that it ideally supports reference speech of arbitrary lengths.
Compared to autoregressive speech language models that prefixes the semantic content by acoustic prompts and models the whole sequence by self-attention only, 
SEF-VC does not suffer from inference stability and speed issues because of the position-agnostic cross-attention method in a non-autogressive manner, either.

\vspace{-0.4cm}
\subsection{Training and Inference}
\label{sybsec:train}
As indicated by Fig.~\ref{fig:model}, we use non-parallel data to train our model. 
Then, a single utterance is divided into two segments. 
The first segment is used to extract the mel-spectrogram providing speaker information, which is cropped from a random starting point with length varying between 2 to 3 seconds randomly. 
The rest of the utterance after the first segment serves as the second, which is fed to the pretrained HuBERT model to extract semantic tokens.
This strategy ensures that the two segments always belong to the same speaker, without the need for oracle speaker labels.
For the auxiliary feature adaptor, during training, we use the output of the first semantic encoder to predict PPE. The ground truth PPE is then added to the output of the first semantic encoder assisting the rest modules for reconstructing waveforms.


The generator loss $\mathcal{L}_G$ is formulated as a weighted sum:
\begin{multline}
\mathcal{L}_G = \lambda_{\mathrm{rec}}\mathcal{L}_{\mathrm{rec}} + \lambda_{\mathrm{feat}}\mathcal{L}_{\mathrm{feat}} + \\
\lambda_{\mathrm{mel}}\mathcal{L}_{\mathrm{mel}} + \lambda_{\mathrm{aux}}\mathcal{L}_{\mathrm{aux}} + \lambda_{\mathrm{adv}}\mathcal{L}_{\mathrm{adv}}
\end{multline}
where $\mathcal{L}_\mathrm{rec}$ is the reconstruction loss measured by $L_1$ distance between mel-spetrograms of real and synthetic waveforms. $\mathcal{L}_{\mathrm{feat}}$ is $L_1$ feature matching loss of intermediate output of the discriminator. $\mathcal{L}_{\mathrm{mel}}$ is measured between the output of the second semantic encoder and the target mel-spectrogram in $L_1$ form. $\mathcal{L}_{\mathrm{aux}}$ is calculated between ground truth $PPE$ and the output $\hat{PPE}$ of auxiliary feature adaptor in $L_1$ form. $\mathcal{L}_{\mathrm{adv}}$ is the adversarial loss in $L_2$ form.

During the inference stage for voice conversion, the target reference speech is used to extract the mel-spectrogram containing speaker information, and source speech is used to obtain semantic tokens through the same pretrained HuBERT model as in the training process.
The PPE is predicted based on the semantic information and reference speech together for the target speaker.
\label{subsec:inference}
\vspace{-0.3cm}
\section{Experiments and Results}
\label{sec:exp}
\subsection{Data and Implementation Details}
\label{subsec:datasets}
Our experiments were conducted on LibriTTS~\cite{zen2019libritts}, a multi-speaker English dataset with a total duration of 586 hours.
We downsampled the utterances to 16kHz and excluded utterances that were too long or too short for training. 
It contains a total of 2456 speakers, including 2311 in the training set, 73 in the validation set, and 72 in the test set. 
To evaluate the model's performance in zero-shot voice conversion, we selected 20 speakers from LibriTTS's test-clean set. Among them, 10 speakers served as source speakers, and 2 utterances of each speaker were selected as the source speeches. 
The remaining 10 served as target speakers and 1 utterance of each speaker with a length of about 3 seconds was selected as the reference speech.

We extracted 1024-dimension semantic features from a pretrained HuBERT model\footnote{\url{https://github.com/facebookresearch/fairseq/blob/main/examples/hubert}} trained on 60k hours of LibriLight. 
Then the semantic features were quantized by K-Means clustering with 2,000 centers offline.
Both semantic encoders consisted of two Conformer blocks, in which multi-head attention layers for self-attention and cross-attention had 2 heads with an attention dimension of 184. 
The mel-encoder contained 1 layer of convolution with kernel size 5 and an output dimension of 184. 
Both generator and discriminator were optimized by Adam with an initial learning rate of 0.0002, and $\beta_1=0.5$, $\beta_2=0.9$, respectively.
The learning rate decreased by 0.5 for every 200k steps. The coefficients of each loss were  $\lambda_{\mathrm{mel}} = 60, \lambda_{\mathrm{aux}} = 5, \lambda_{\mathrm{rec}} = 45, \lambda_{\mathrm{adv}} = 1, \lambda_{\mathrm{feat}} = 2$. 
Besides, the frameshift of reference mel-spectrograms was 10ms to better capture acoustic details, especially speaker identity, while the frame shift of semantic tokens was 20ms.
\vspace{-0.3cm}
\subsection{Baselines}
We compared SEF-VC with the following zero-shot VC methods:
\begin{itemize}[leftmargin=*,noitemsep, topsep=5pt]
    \item \textbf{AdaIN-VC}~\cite{adain-vc} that disentangles speaker and content information by simply introducing instance normalization to the auto-encoder bottleneck. It relies on speaker embeddings for speaker representation modeling.
    \item \textbf{YourTTS}~\cite{casanova2022yourtts} that performs speaker disentanglement by normalizing flows. It conducts VC by separating source speaker information in the reversed flow process, and plug the target speaker embedding in the forward flow.
    \item \textbf{Polyak et al.}~\cite{polyak2021speech} that separates speaker and semantic information with a disentangled representation of speech content, prosody, and speaker identity information. It uses pretrained encoders to extract semantic tokens, pitch tokens, and target speaker embeddings respectively, and synthesizes waveform with a HifiGAN generator. 
    For ease of presentation, we refer to it as \textbf{SSR-VC} hereafter.
\end{itemize}
For fair comparison, we trained and tested AdaIN-VC\footnote{\url{https://github.com/jjery2243542/adaptive_voice_conversion}}, YourTTS\footnote{\url{https://github.com/Edresson/YourTTS}} and SSR-VC\footnote{An unofficial implementation by us.} on the same data partition with SEF-VC in LibriTTS.

\vspace{-0.3cm}
\subsection{Any-to-Any Voice Conversion Results}
\label{subsec:vc results}
We conducted objective and subjective evaluations to evaluate the speaker similarity and speech intelligibility of SEF-VC on any-to-any voice conversion. 
The objective evaluations included speaker embedding cosine similarity (SECS) and character error rate (CER) in ASR.
The SECS metric was computed by extracting speaker embeddings with Resemblyzer\footnote{\url{https://github.com/resemble-ai/Resemblyzer}} and calculating the cosine similarity.
The CER was measured between transcripts of synthesized and real utterances transcribed by an ASR model\footnote{\url{https://catalog.ngc.nvidia.com/orgs/nvidia/teams/nemo/models/stt\_en\_quartznet15x5}} following ~\cite{hussain2023ace}.
We also performed subjective evaluations by mean opinion score (MOS) tests to measure speaker similarity, where raters were asked to score by how similar the synthesized voice was to the reference on a scale of 1-5.
Naturalness MOS tests were also conducted to measure the intelligibility of synthesized voice.
The SECS, CER, similarity, and naturalness MOS results can be seen in Table \ref{tab:main}. 

The results demonstrate our model can convert voice to target speaker better in any-to-any voice conversion. AdaIN-VC is limited by the bottleneck in auto-encoder, and thus synthesizes voice with poor quality. 
YourTTS and SSR-VC both heavily depend on global speaker embedding, which lacks enough target speaker information, resulting in suboptimal conversion performance. 
This drawback of global speaker embedding will also be verified in section \ref{subsub:abla spk}.
On the contrary, the proposed speaker embedding free SEF-VC using  position-agnostic cross-attention can better capture and incorporate speaker information, hence obtaining better speaker similarity.

\begin{table}[ht]
\centering
\small{
\begin{tabular}{@{}c|cc|cc@{}}
\toprule
\multirow{2}{*}{} & \multirow{2}{*}{SECS$\uparrow$} & \multirow{2}{*}{CER$\downarrow$} & \multicolumn{2}{c}{MOS}  \\ \cmidrule(l){4-5} 
 & & & Similarity & Naturalness \\ \midrule
AdaIN-VC~\cite{adain-vc}           &0.768           &15.7\%  &3.20±0.09            &3.15±0.07   \\
YourTTS~\cite{casanova2022yourtts} &0.770           &13.6\%         
& 3.55±0.09           &3.56±0.09\\
SSR-VC~\cite{polyak2021speech}     &0.804           &\textbf{5.52\%} & 4.22±0.12           &4.16±0.11        \\
SEF-VC                             &\textbf{0.825}  &5.53\%         
& \textbf{4.52±0.08}  &\textbf{4.37±0.09} \\ \midrule
GT & - &4.80\% & - & 4.56±0.09 \\
\bottomrule
\end{tabular}%
}
\vspace{-0.2cm}
\caption{Performance comparison between SEF-VC and other baselines in any-to-any voice conversion. SECS means speaker embedding cosine similarity. Higher SECS means higher speaker similarity while lower CER means more intelligible.}
\label{tab:main}
\end{table}
\vspace{-0.5cm}

\subsection{Different Reference Lengths}
\label{subsub:abla pro}
\begin{figure}[ht]
    \centering
    \includegraphics[width=0.99\linewidth]{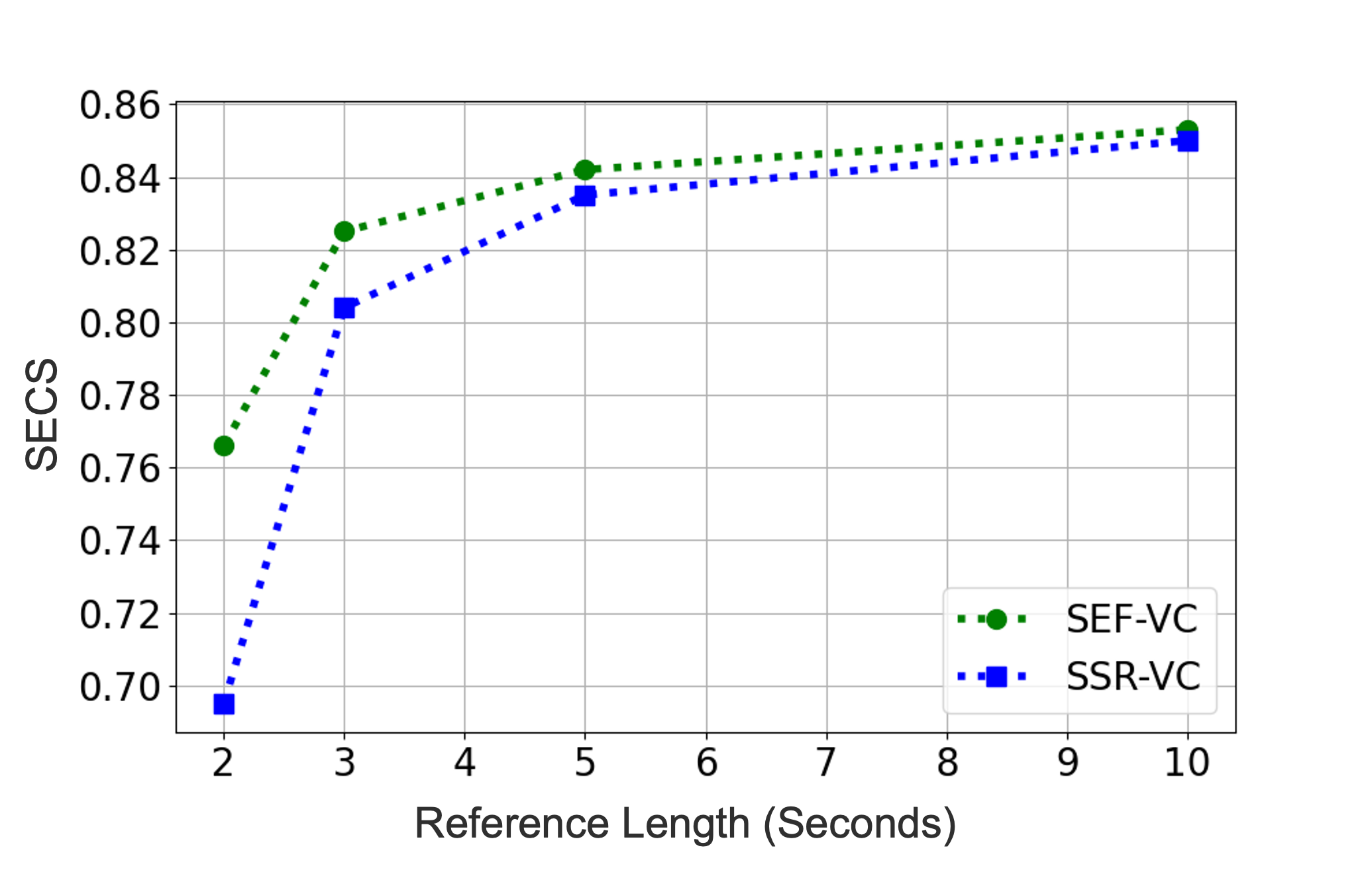}
    \vspace{-0.5cm}
    \caption{Experiment on the influence of references of different lengths on zero-shot VC.}
    \label{fig:abal prompt}
\end{figure}

In this section, we inspect the impact of different lengths of prompts within SEF-VC and SSR-VC~\cite{polyak2021speech}, including 2s, 3s, 5s, and 10s. The results can be seen in Figure \ref{fig:abal prompt}. In SSR-VC~\cite{polyak2021speech}, the prompt serves to extract speaker embedding. 
The results show that the performance of our model improves with the length of reference speech getting longer, which benefits from our proposed position-agnostic cross-attention mechanism. 
Intuitively, a longer prompt makes it easier to model the speaker information, thus more helpful to convert the voice to target speakers. 
Even in the case where the reference speech length is as short as 2s, the SECS of SEF-VC is still acceptable compared to SSR-VC~\cite{polyak2021speech}, which shows cross-attention mechanism is more robust than speaker embedding.
Starting from a reference length of 3s, SEF-VC can already capture the target speaker's timbre to a great extent.

\vspace{-0.4cm}

\subsection{Cross-Attention vs Speaker Embedding}
\label{subsub:abla spk}
In this part, we demonstrate the effectiveness of our proposed cross-attention mechanism. For comparison, we remove the cross-attention module in SEF-VC and add speaker embedding directly into the  HifiGAN~\cite{kong2020hifi} generator following YourTTS~\cite{casanova2022yourtts}. X-vector~\cite{snyder2018x} is regarded as speaker embedding extracted by Kaldi~\cite{povey2011kaldi}. 
The result is shown in table \ref{tab:abal spk}.
It can thus be seen that the adopted cross-attention mechanism greatly improved the speaker similarity, which means it can better learn and incorporate target speaker information with latent content representation than simple speaker embedding addition. 
One possible reason is that global speaker embedding cannot provide enough speaker-dependent information, like time-varying pitch, which contributes to speaker similarity too.
\begin{table}[ht]
\centering
\small{
\begin{tabular}{@{}c|ll|ll@{}}
\toprule
\multirow{2}{*}{Speaker Modeling } & \multirow{2}{*}{SECS$\uparrow$} & \multirow{2}{*}{CER$\downarrow$} & \multicolumn{2}{c}{MOS}  \\ \cmidrule(l){4-5} 
& & & Similarity & Naturalness \\ \midrule
speaker embedding &0.711    &\textbf{5.38\%}     &3.84±0.13  &4.29±0.09     \\
cross-attention   &\textbf{0.825}     &{5.53\%}     &\textbf{4.52±0.08}   &\textbf{4.37±0.09}    \\
\midrule
GT      &     -      &4.80\%     &    -        &4.56±0.09     \\
\bottomrule
\end{tabular}%
}
\vspace{-0.2cm}
\caption{Experiment on the effectiveness of different speaker modeling.}
\label{tab:abal spk}
\end{table}

\vspace{-0.5cm}
\section{Conclusion}
\label{sec:conclusion}
We proposed SEF-VC, a speaker embedding free voice conversion model, which is designed to learn and incorporate speaker timbre from reference speech via a position-agnostic cross-attention mechanism, and then reconstruct waveform from HuBERT semantic tokens in a non-autoregressive manner. The concise design of SEF-VC enhances its training stability and voice conversion performance. Subjective and objective evaluations show that our model can generate natural-sounding speech similar to the target speaker. Ablation studies further demonstrate the effectiveness of the position-agnostic cross-attention mechanism, which allows as short as 2-second reference speech for voice conversion.

\vspace{-0.3cm}
\section{Acknowledgement}
This work was supported by China NSFC Project (No. 92370206), Shanghai Municipal Science and Technology Major Project (2021SHZDZX0102) and the Key Research and Development Program of Jiangsu Province, China (No. BE2022059).

\vfill\pagebreak

\bibliographystyle{IEEEtran}
\small{
\bibliography{refs}
}

\end{document}